\documentclass[aps,prb,reprint,groupaddress,amsmath,amssymb]{revtex4-1}
\usepackage{epsfig,psfrag}
\usepackage{dcolumn}
\usepackage{bm}
\usepackage{graphicx}
\usepackage{amsmath}
\usepackage{color}

\newcommand{\MF}{}
\newcommand{\E}{\mathbf{E}}
\renewcommand{\H}{\mathbf{H}}
\newcommand{\Ham}{\mathcal{H}}
\renewcommand{\P}{\mathbf{P}}
\newcommand{\B}{\mathbf{B}}
\newcommand{\D}{\mathbf{D}}
\newcommand{\A}{\mathbf{A}}

\newcommand{\q}{\bm{q}}
\newcommand{\rr}{\bm{r}}
\newcommand{\inv}{${}^{-1}$}

\begin{document}
\preprint{}
\title{Quantum model of gain in phonon-polariton lasers}

\author{M. Francki\'e}
\email{Electronic mail: martin.franckie@phys.ethz.ch}
\author{C. Ndebeka-Bandou}
\author{K. Ohtani}
\author{J. Faist}
\affiliation{Institute for Quantum Electronics, ETH Zurich, Auguste-Piccard-Hof 1, 8093 Zurich, Switzerland} 

\begin{abstract}
We develop a quantum model for the calculation of the gain of phonon-polariton intersubband lasers. The polaritonic gain \MF{arises} from the interaction between electrons confined in a quantum well structure and phonons confined in one layer of the material. Our theoretical approach is based on expressing the \MF{resonant} matter excitations (intersubband electrons and phonons)  in terms of polarization densities in second quantization, and treating all non-resonant polarizations with an effective dielectric function. The interaction between the electronic and phononic polarizations is treated perturbatively, and gives rise to stimulated emission of polartions in the case of inverted subbands. Our model provides a complete physical insight of the system and allows to determine the phonon and photon fraction of the laser gain. Moreover, it can be applied and extended to any type of designs and material systems, offering a wide set of possibilities for the optimization of future phonon-polariton lasers. 
\end{abstract}

\pacs{73.21.Ac,78.67.Pt}

\maketitle

%
%
\section{Introduction}
A polariton~\cite{hopfield_theory_1958} is a composite excitation arising from the coupling of light with a material excitation. As such, polaritons are exhibiting properties that are inherited from their two original constituents and can be tailored over a large range through the strength of \MF{the} light-matter coupling. Polaritons can be seen as forced to interact through their matter part, while the photons will carry the imprint of the coherence properties, enabling their measurement in the far-field using photodetectors. For well-chosen experimental parameters, the polaritons exhibit features of a quantum fluid whose properties have attracted a lot of attention recently.\cite{carusotto_quantum_2013}

While much attention has been given to exciton-polaritons and their properties in the visible,\cite{hopfield_theory_1958,deng_condensation_2002,kasprzak_boseeinstein_2006,deng_exciton-polariton_2010,berini_surface_2012} the study of polaritons in the mid-infrared portion of the spectrum has also some unique features.\cite{colombelli_quantum_2005,huber_near-field_2005,korobkin_mid-infrared_2007} The coupling between an intersubband electronic system and longitudinal optical (LO) phonons was described recently as an intersubband polaron,~\cite{de_liberato_quantum_2012} and the coupling between an intersubband system and light, called an intersubband (or cavity) polariton,\cite{dini_microcavity_2003,ciuti_quantum_2005} was theoretically investigated in the Power-Zienau-Woolley (PZW) gauge\cite{power_radiative_1957,woolley_molecular_1971} by Todorov \emph{et al.} in Ref.~\onlinecite{todorov_intersubband_2012}. In addition, the light can resonantly couple to transverse optical (TO) phonons forming phonon-polaritons,\cite{hopfield_aspects_1995} which have mostly been studied at the surface of polarizable materials,\cite{gubbin_real-space_2016,christopher_r._gubbin_theoretical_2017} and recently also in the bulk using classical theory.\cite{dzedolik_phonon-polaritons_2016,gomez-urrea_light_2017,ramanujam_effect_2017} The strong coupling properties have also been observed experimentally and described using a dielectric function approach.\cite{askenazi_ultra-strong_2014}

\begin{figure}
 \includegraphics[width=0.8\linewidth]{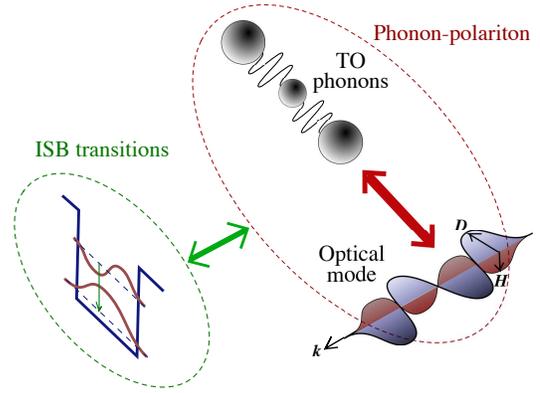}
 \caption{(Color online) Scheme of the three interactions involved in the lasing process of a phonon-polariton laser. The red arrow symbolizes the strong coupling between the cavity modes and the TO phonon modes that creates the phonon-polariton. The green arrow symbolizes the weak interaction between the phonon-polariton and the ISB transitions that generates the laser gain. }
 \label{fig_intro}
\end{figure}

As shown schematically in Fig.~\ref{fig_intro}, when the cavity {\em and} \MF{the} intersubband transitions are chosen to be energetically resonant with a mechanical resonance of the semiconductor lattice, a unique tripartite coupling can be achieved.  For large electron concentrations and in thermal equilibrium, the resulting polaritonic dispersion arises due to the coupling of light to both excitations.
An interesting feature of the intersubband system is that it can be electrically excited, providing optical gain. Solid state phonon lasers were proposed \cite{wolff_plasma-wave_1970} and analysed using either a pure phononic gain \cite{chen_feasibility_2003} or using an electronic Raman approach.\cite{khurgin_stimulated_2006}
 In contrast, we present a fully quantized model that treat both the photons and phonons, as well as the inter-subband system on an equal footing, in the PZW gauge. This allows us to account for the spatial variation of the material optical response, as the phonon polarization
 is spatially confined. We thus fully account for the tripartite coupling, albeit using a basis of phonon-polaritons, since the photon-phonon coupling is the stronger one. In this basis, the rate of stimulated emission of phonon-polaritons from inter-subband excitations, is derived in first order perturbation theory. We also provide computational examples for a resonant tunnelling diode (RTD) and a quantum cascade laser (QCL), where the phonon-polaritons are confined to potential barriers in the conduction band profile. However, the theory can easily be expanded to account for arbitrary 2D heterostructures, as well as other material excitations provided their quantized polarization.

This paper is organized as follows: In section \ref{sec:class} we derive the classical Hamiltonian for oscillating polarization densities in the presence of a time-dependent electro-magnetic field as the starting point of our quantum formulation. Then, we quantize this Hamiltonian in section \ref{sec:quant} by introducing the polarization density operators for the intersubband system (Sec.~\ref{sec:elpol}) and the relevant phonon excitations (Sec.~\ref{sec:phonpol}), in second quantization. In Sec.~\ref{sec:polariton}, we derive the interaction Hamiltonian for the phonon and photon fields, which is then diagonalized to give the phonon-polariton creation-annihilation operators and dispersion relation. Finally, in Sec.~\ref{sec:polisb} we describe the polariton-ISB interaction responsible for the polartion gain and provide computational examples of the model in Sec.~\ref{sec:examples}.

\section{Classical formulation}\label{sec:class}
The starting point for our model is the Lagrangian density \cite{kittel_quantum_1963}
\begin{eqnarray}
{\cal L} &=& \frac{\varepsilon_0}{2} \left(\dot{\A} + \nabla\phi\right)^2 - \frac{1}{2\mu_0} \left(\nabla\times \A\right)^2\nonumber \\ 
 &+& \sum_i\left(\frac{1}{2\chi_i}\dot{\P}_i^2 - \frac{\omega_i^2}{2\chi_i}\P_i^2 -\P_i(\dot{\A} + \nabla\phi)\right)\label{eq:Lagrangian}
\end{eqnarray}
for a polarization $\P_i$ in vacuum, represented by a sum of harmonic oscillators with eigenfrequency $\omega_i$ and "mass" $\chi_i$, under the application of an electro-magnetic field with vector and scalar potentials $\A$ and $\phi$ . The last term accounts for the electric potential energy stored in the "springs" of the oscillators. 
This leads to the Hamiltonian
\begin{eqnarray}
\Ham &=& \int d^3r \Big{[} \frac{1}{2\varepsilon_0}\D^2 + \frac{1}{2\mu_0} \left(\nabla\times \A\right)^2 + \frac{1}{2}\sum_{i\neq j}\P_i\P_j \label{eq:EqHKittel2} \\&+& \sum_i\left( \underbrace{\left(\frac{\omega_i^2}{2\chi_i}+\frac{1}{2\varepsilon_0} \right)}_{\omega_i'/2\chi_i}\P_i^2 + \frac{1}{2\chi_i}\dot{\P}_i^2- \frac{1}{\varepsilon_0}\D\cdot \P_i\right)\Big{]}\nonumber,
\end{eqnarray}
where $\D = -\varepsilon_0(\dot{\A} + \nabla\phi)+\sum_i \P_i$ is the electric displacement field satisfying $\nabla\cdot\D = 0$.
Here, the frequencies $(\omega_i')^2 = \omega_i^2 + \omega_{P,i}^2$ of the oscillators are shifted by the plasma frequency $\omega_{P,i}^2 = \frac{\chi_i}{\varepsilon_0}$  with respect to the bare mechanical frequency of Eq.~\eqref{eq:Lagrangian}.
We shall consider only a few polarization terms, namely those coming from TO phonons and the electrons confined in the conduction band. We thus treat all non-resonant oscillators with an effective dielectric constant defined by
\begin{equation}
\D = \varepsilon_0\E + \sum_{i\in  \text{b.}} \P_i + \sum_{i\in  \text{e}, \MF{L}} \P_i\equiv \varepsilon_r\varepsilon_0 \E + \sum_{i\in  \text{e}, \MF{L}} \P_i\,
\end{equation}
where the first sum is over the background (b), and the second over the ISB (e) and the resonant \MF{lattice ($L$)} polarizations. Assuming $\P_{i\in\text{b}}$ oscillate at the cavity frequency $\omega$,
\begin{equation}
\varepsilon_r = 1 + \sum_{i\in b} \frac{\chi_i}{\varepsilon_0(\omega_i^2 - \omega^2)}.
\end{equation}
Considering only the background, the Hamiltonian reads
\begin{eqnarray}
\Ham_\text{b} &=& \int d^3r \left(\frac{1}{2}\E\cdot \D + \frac{1}{2} \B\cdot\H \right) + \Ham_\text{\MF{mat}} \\
&=& \int d^3r \left(\frac{1}{2\varepsilon_0\varepsilon_r(z)} \D^2 + \frac{\mu_0}{2}\H^2 \right)+ \Ham_\text{\MF{mat}},
\end{eqnarray}
where $\nabla\times\A = \B = \mu_0\H$ and we assumed in Eq.~\eqref{eq:Lagrangian} that there are no magnetic moments in the system. $\Ham_\text{\MF{mat}}$ contains all terms of Eq.~\eqref{eq:EqHKittel2} which \MF{contain the background matter polarizations $P_i$ only}.
Physically, this term contains the energy contribution of all the crystal ions, and will not affect the following theory where the we treat the conduction band electrons in the envelope function approximation. We will thus suppress this term from now on.
Adding the special polarizations $\sum_{i\in\text{e}, \MF{L}} \P_i$, not included in $\varepsilon_r$, we find
\begin{eqnarray}
\Ham &=& \int d^3r \left(\frac{1}{2\varepsilon_0\varepsilon_r}\D^2 + \frac{\mu_0}{2}\H^2\right) + \nonumber  \\ &+& \frac{1}{\varepsilon_0\varepsilon_r}\int d^3 r \left( - \sum_{i\in\text{e}, \MF{L}}\D\cdot\P_i + \frac{1}{2}\sum_{(ij)\in\text{e}, \MF{L}}\P_i\P_j\right) \nonumber\\ 
&+& \int d^3 r \sum_{i\in{e}, \MF{L}} \frac{1}{2\chi_i} \left( \omega_i^2 \P_i^2 + \dot{\P}_i^2 \right).\label{eq:hclass}
\end{eqnarray}
This Hamiltonian resembles the one of Ref.~\onlinecite{todorov_intersubband_2012}. We will use Eq.~\eqref{eq:hclass} and diagonalize the terms in the quantized Hamiltonian containing $\P_{\MF{L}}$ only in Sec.~\ref{sec:polariton}. We also note that in a heterostructure, $\varepsilon_r = \varepsilon_r(z,\omega)$ will acquire a $z$-dependence which in principle has to be considered when performing the volume integral.

\MF{The background dielectric function $\varepsilon_r$ results from both inter-atomic polarizations and bound electrons. In addition, electrons in quantum states spatially separated from the resonant phonon polarization may give a small contribution. We will assume that $\varepsilon_r$ is close to the bulk values of the constituent materials, why we will later set $\varepsilon_r=\varepsilon_{\infty}$ of the bulk well material.}

\section{Quantum formulation} \label{sec:quant}
In order to quantize the system, we write down the quantized version of the Hamiltonian \eqref{eq:hclass} as
\begin{equation}
\hat{\mathcal{H}} = \hat{\mathcal{H}}_{rad} +  \MF{\hat{\mathcal{H}}_{L} + \hat{\mathcal{H}}_{e}} + \hat{\mathcal{H}}_{int},
 \label{eq_general_hamiltonian}
\end{equation}
where the radiation in the cavity is \MF{written in second quantization} as
\begin{equation} 
\hat{\mathcal{H}}_{rad} = \sum_{\bm{q}}\hbar\omega_{\mathrm{cav},\bm{q}}\left( a_{\bm{q}}^\dagger a_{\bm{q}}+\frac{1}{2} \right).
\end{equation}
For the cavity modes, we use the quantized displacement field of the TM mode in \MF{the PZW gauge (see e.~g.~Ref.~\onlinecite{cohen-tannoudji_photons_1989})}
\begin{equation}
\hat{D}_z(R) = i\sum_{\q} \sqrt{\frac{\varepsilon_r \varepsilon_0\hbar\omega_{\text{opt},\q}}{2SL_\text{cav}}}e^{iq\cdot r}g_q(z)(a_q - a^\dagger_{-q}),
\end{equation}
where $S$ and $L_\text{cav.}$ is the surface area and length of the cavity, repsectively, and $g_q(z)$ is the mode profile normalized as
\begin{equation}
\int_{-\infty}^{\infty} g_q^2(z) dz = L_\text{cav.}.
\end{equation} 
These are the only modes than can propagate in 2D heterostructures.
Still neglecting magnetic interactions, the light-matter interaction in Eq.~\eqref{eq:hclass} leads to an interaction Hamiltonian having the form
\begin{equation}
\hat{\mathcal{H}}_{int} =\int d^3 r  \frac{1}{\varepsilon_0 \varepsilon_r }  \left ( -\hat{{\bf D}} \cdot {\hat {\bf P}}_{mat} + \frac{1}{2} \hat{{\bf P}}_{mat}^2 \right ),
\end{equation}
where the sum in Eq.~\eqref{eq:hclass} runs over the intersubband transitions and the lattice contributions $\hat{{\bf P}}_{mat} = \hat{{\bf P}}_{e} + \hat{{\bf P}}_{L}$, respectively.

The formalism developed so far does not take into account any dissipative couplings for the phonons, electrons, or photons. In the first part of this paper, we will completely neglect these coupling terms. Later, when we calculate the polariton gain, however,
we need to include the phonon and photon decays via an effective decay rate into acoustic \MF{phonons} and cavity losses, respectively. For the ISB system, dissipation due to optical and acoustic phonons, as well as elastic scattering with ion impurities, alloy disorder, and interface roughness, can be included in the transport calculations providing the self-consistent populations of the ISB levels undergoing stimulated polariton emission.

Since we are interested in a situation where the electronic system provides optical gain but will remain in the weak coupling with radiation, we split the Hamiltonian
\begin{align}
\hat{\mathcal{H}} = & \underbrace{\hat{\mathcal{H}}_{rad} + \hat{\mathcal{H}}_{L} +  \int d^3 r \frac{1}{\varepsilon_0 \varepsilon_r }\left ( -\hat{{\bf D}} \cdot {\hat {\bf P}}_{L}    + \frac{1}{2} \hat{{\bf P}}_{L}^2 \right )}_{\textrm{Diagonalize } \hat{\mathcal{H}}_P} \nonumber \\ 
&+ \underbrace{\int d^3 r \frac{1}{\varepsilon_0 \varepsilon_r } \left ( -\hat{{\bf D}} \cdot {\hat {\bf P}}_{e} + \frac{1}{2}{\hat{\bf P}}_{L}\cdot{\hat{\bf P}}_{e} \right ) }_{\textrm{Perturbation}} \nonumber \\
&+ \underbrace{\int d^3 r \frac{1}{\varepsilon_0 \varepsilon_r } \frac{1}{2} \hat{{\bf P}}_{e}^2}_{\textrm{Neglect}}  + \hat{\mathcal{H}}_{e}
\end{align}
 into three parts. The first part $\hat{\mathcal{H}}_{P}$ contains the lattice-radiation coupling and will lead to our polaritonic basis after diagonalization. Amplification or attenuation of these polaritons through their interaction with the intersubband system will be computed using Fermi's golden rule applied to the second part of \MF{$\hat{\mathcal{H}}$}. Finally, as we are dealing with a low electron population, we can safely neglect the intersubband polarization self-energy, which accounts for the depolarization shift.

\section{Electron polarization}\label{sec:elpol}

The electronic subband states $n$ are defined by their energies $E_n(\bm{k})=E_n + \frac{\hbar^2 \bm{k}^2}{2m^*}$ and their wavefunctions $\langle \bm{r},z|n,\bm{k}\rangle=\frac{1}{\sqrt{S}}e^{i\bm{k}\cdot\bm{r}}\chi_n(z)$,  where $S$ is the sample area, $m^*$ is the carrier effective mass, $\bm{k}$ and $\bm{r}=(x,y)$ are the in-plane wavevector and the in-plane coordinate respectively. 
Starting from the initial level $|i,\bm{k}\rangle$ (where the carriers are either residing or electrically injected), each possible ISB transition $|i,\bm{k} \rangle \rightarrow |n,\bm{k'} \rangle$ is labeled by the index $j$ and occurs at a frequency $\omega_{j}=\omega_{i}-\omega_{n}$.   The intersubband polarization is\cite{todorov_intersubband_2012}
\begin{equation}
\hat{{\bf P}}_{e} ({\bf r}) = \frac{\hbar e }{2 S m^*} \sum_{j,\bm{q}} \frac{\sqrt{\Delta N_j}\xi_{j}(z)}{\omega_{j}}  e^{i \bm{q} \bm{r}} \left [ b_{j,-\bm{q}}^\dagger + b_{j,\bm{q}} \right ],
\end{equation}
where $e = -|e|$ is the electron charge, $\Delta N_j = N_i-N_n$ is the population inversion, and is expressed as a function of the bright mode creation operators (assuming that the transitions are vertical in the  $\bm{k}$-space ($\bm{k}\approx\bm{k'}$))
\begin{equation}
 b^{\dagger}_{j, \bm{q}} = \frac{1}{\sqrt{\Delta N_j}}\sum_{{\bf k}} c^{\dagger}_{n,\bm{k}} c_{i,\bm{k}} \equiv b^\dagger_j.
 \end{equation}
 \MF{Here, $c^{\dagger}_{n,\bm{k}}$ and  $c_{i,\bm{k}}$ are the creation and anihilation operators for the one-electron ISB states $|n,\mathbf{k}\rangle$.}
The microcurrents for the transition between state are defined from the wavefunctions as 
\begin{equation}
\xi_j(z)=\chi_{i}(z)\partial_z\chi_n(z)-\chi_n(z)\partial_z\chi_{i}(z).
\end{equation}
In this formalism, the electron Hamiltonian is expressed as 
\begin{equation}
\hat{\mathcal{H}}_\text{e} = \sum_{j}b^\dagger_jb_j\hbar\omega_j.
\end{equation}

\section{Phonon polarization} \label{sec:phonpol}
The only lattice vibrations interacting with a light field are the transverse optical (TO) phonons, which are assumed to be dispersionless and have a mechanical frequency $\omega_{\text{TO}}$. We will assume that the vibrations are localized in \MF{layers in the $x$-$y$ plane, with $z$-coordinate} $z_i$. The phonon polarization (see Appendix \ref{sec:phonapp})
\begin{equation}
\hat{P}_{L,z}(R) =  \frac{\hbar e}{2SM}\sum_{\q,i}\frac{\xi_{\MF{L},i}(z)}{\omega_\text{TO}} e^{i\q\cdot \rr}(d^\dagger_{i,-\q}+ d_{i,\q}).\label{eq:phonpol}
\end{equation}
is similar to the electronic one, with $M$ being the vibrational mass and $d^\dagger_{i,-\q}$ creating a TO phonon with in-plane momentum $-\q$. The phonon micro current is (see the appendix)
\begin{equation}
\xi_{\MF{L},i} = \sqrt{\frac{2}{\pi}}\frac{1}{\sigma^2}e^{-(z-z_i)^2/\sigma^2}
\end{equation}
and $\sigma$ \MF{is related to the standard} deviation from the equilibrium position $z_i$.
These phonons are represented globally by the Hamiltonian
\begin{equation}
\hat{\mathcal{H}}_{L} = \sum_{\q i}\hbar \omega_{\text{TO}} d_{\q i}^\dagger d_{\q i}, \label{eq:Hphon}
\end{equation}
provided the phonon plasma frequency is
\begin{equation}
\omega_{P,i}^2 = \frac{\hbar e^2 \int \xi_{\MF{L},i}^2(z) \text{d}z}{2SM^2\epsilon_0\epsilon_r\omega_\text{TO}}.\label{eq:plasmaf}
\end{equation}
The operator $d_{i,\q}^\dagger$ can be thought of as creating a phonon excitation confined in one monolayer, traveling with momentum $\q$ inside this layer. The polarization operator \eqref{eq:phonpol} depends on the density of oscillators via the plasma frequency, which fixes the oscillator "mass" $M$. 
The index $i$ can either represents physical atomic layers, or spatially separated thin layers of the bulk material (so thin that the lattice ions oscillate in phase and the phonon micro current can be represented by a gaussian function).

\section{Hamiltonian of the phonon-polariton}\label{sec:polariton}
The diagonalization of $\hat{\mathcal{H}}_P$ in thermal equilibrium, assuming all the carriers are in the ground state, yields polaritons that combine lattice and electronic excitations,\cite{hopfield_theory_1958} and have been observed experimentally.\cite{askenazi_ultra-strong_2014}
In this section, we will diagonalize $\hat{\mathcal{H}}_{P}$ in two steps and find the polariton eigenstates. First, we will incorporate the phonon polarization self-energy $\frac{1}{2\varepsilon_0\varepsilon_r}\hat{{\bf P}}_{L}^2$ into the bare phonon Hamiltonian $\hat{\mathcal{H}}_{L}$, which will lead to an energy renormalization similar to the depolarization shift of the intersubband system. The following calculation will be significantly lightened by neglecting the terms of $\P_L^2$ mixing $\xi_{P,i}$ with different layer indices $i$, as motivated in 
Appendix \ref{sec:phonapp}.
Thus, we diagonalize
\begin{eqnarray}
&&\hat{\mathcal{H}}'_L \equiv \hat{\mathcal{H}}_{L} + \int d^3 r \frac{1}{2\varepsilon_0 \epsilon_r} {\hat {\bf P}}_{L}^2 = \hbar\omega_\text{TO} \sum_{i\q} d^\dagger_{i,\q}d_{i,\q} +\nonumber \\
&& \sum_{i\q} \hbar\Theta_i (d^\dagger_{i,\q} + d_{i,-\q})(d^\dagger_{i,-\q} + d_{i,\q}),
\label{eq:Hphot}
\end{eqnarray}
where
\begin{equation}
\Theta_i= \frac{e^2\hbar}{8SM^2\varepsilon_0\epsilon_r \omega_\text{TO}^2} \int \xi^2_{\MF{L},i}(z) dz,\label{coupling_great_theta}
\end{equation}
(here, we assumed that the background dielectric constant does not vary on the scale of the phonon layers), with the new operators
\begin{eqnarray}
p_{i,\q} = \frac{\omega'_i - \omega_\text{TO}}{2\sqrt{\omega'_i\omega_\text{TO}}} d^\dagger_{i,-\q}+
\frac{\omega'_i + \omega_\text{TO}}{2\sqrt{\omega'_i\omega_\text{TO}}} d_{i,\q}.
\end{eqnarray}
The eigenvalue is $(\omega'_i)^2 = (\omega_\text{TO})^2 + 4\omega_\text{TO}\Theta_i$, and interpreting 
this as the longitudinal optical (LO) phonon frequency, we deduce the plasma frequency $\omega_{P,i}^2 = 4\omega_\text{TO}\Theta_i$.
%
Expressing $\hat{\mathcal{H}}_{P}$
in second quantization format now leads to the expression
\begin{align}
  \hat{\mathcal{H}}_P=&\sum_{\bm{q}}\hbar\omega_{\mathrm{cav},\bm{q}}\left( a_{\bm{q}}^\dagger a_{\bm{q}}+\frac{1}{2} \right) 
 +\sum_{\q,i}\hbar\omega'_ip_{i,\q}^\dagger p_{i,\q}\nonumber\\
 &+i\sum_{\q,i}\hbar\Lambda_{i,\q}\left( a_{\q}^\dagger-a_{-\q}\right)\left( p_{i,-\q}^\dagger+p_{i,\q}\right)
 \label{hamiltonian_phonon_photon}
\end{align}
with 
\begin{align}
\label{coupling_great_lambda}
 \Lambda_{i,\q}&=\frac{\omega_{P,i}}{2}\sqrt{\frac{\omega_{\mathrm{opt},\bm{q}}}{\omega'_i}f_{P,i}},
 \end{align}
where 
\begin{equation}
f_{\MF{L},i}=\left(\frac{\int g_{\bm{q}}(z)\xi_{\MF{L},i}(z)dz}{\sqrt{L_\mathrm{cav.}\int \xi^2_{\MF{L},i}(z)dz}}  \right)^2.
\end{equation} 
The factor $f_{\MF{L},i}$ measures the filling of the cavity by the mechanical oscillators and is equal 1 for the bulk material.

Proceeding with the second step of the diagonalization of $\hat{\mathcal{H}}_P$, the Hamiltonian (\ref{hamiltonian_phonon_photon}) can be exactly diagonalized through a Bogoliubov transformation and by the introduction of the polariton operator $\Pi_{\bm{q}}=x_{\bm{q}}a_{\bm{q}}+y_{\bm{q}}a^\dagger_{\bm{-q}}+z_{\bm{q}}p_{\bm{q}}+t_{\bm{q}}p^\dagger_{\bm{-q}}$.
The two real solutions  $\omega_{\bm{q},\pm}$ of the eigenvalue equation $[\Pi_{\bm{q}},\hat{\mathcal{H}}_\mathrm{P}]=\hbar\omega_{\bm{q}}\Pi_{\bm{q}}$ are the frequencies of the two polaritonic branches and are readily obtained by
\begin{align}
\label{eigenvalue0}
\omega_{\bm{q},\pm}&=\frac{1}{\sqrt{2}}\sqrt{\omega'^2+\omega^2_{\mathrm{opt},\bm{q}}\pm\sqrt{\Delta}}\\
\Delta&= \omega'^4+4\omega^2_\mathrm{c}\omega^2_{\mathrm{opt},\bm{q}}-2\omega'^2\omega^2_{\mathrm{opt},\bm{q}}+\omega^4_{\mathrm{opt},\bm{q}}
\label{eigenvalue}
\end{align}
with $\omega^2_\mathrm{c}= f_p \omega_\mathrm{P}^2$. The two polaritonic branches have the asymptotes $\omega_{\q\rightarrow 0,+}=\sqrt{\omega_\text{TO}^2 + \omega_P^2}\equiv\omega'$ and $\omega_{\q\rightarrow \infty,-}=\sqrt{\omega^2_\text{TO}+\omega^2_P(1-f_P)}\equiv\omega''$. Additionally, the diagonalization of (\ref{hamiltonian_phonon_photon}) in the polaritonic basis allows to determine the mixing fractions of the phonon-polariton, namely its photonic ($h_{l,\bm{q}}=|x_{\bm{q}}|^2-|y_{\bm{q}}|^2$) and phononic ($h_{p,\bm{q}}=|z_{\bm{q}}|^2-|t_{\bm{q}}|^2$) fractions. For instance, in the upper branch ($\omega_{\bm{q}}=\omega_{\bm{q},+}$), we have the fractions
\begin{equation}
  h^+_{l,\bm{q}}=\frac{\omega_{\bm{q},+}^2-\omega_\text{TO}^2}{\omega_{\bm{q},+}^2-\omega_{\bm{q},-}^2},\quad 
 h^+_{p,\bm{q}}=1-h^+_{l,\bm{q}}.
 \label{hopfield_coefficients}
\end{equation}
In the lower branch, the mixing fractions are simply obtained by $h^-_{l,\q} = 1-h^+_{l,\q}$.
While the limits when $\omega_{\bm{q},+}\rightarrow\omega'$ and $\omega_{\bm{q},+}\rightarrow\omega_\mathrm{TO}$ correspond to mostly phonon states, in the vicinity of the anti-crossing the mixing fractions reach a value of $0.5$, indicating a maximum phonon-photon admixture. A suitable design of the active region enables to achieve a lasing emission at frequencies close to this maximum adxmiture point, where a non-vanishing phononic gain is therefore expected. 

If the filling factor $f_{\MF{L}}$ is small (as e.~g.~for a thin-layer structure such at those in Figs.~\ref{fig:RTD} and \ref{fig:QCL}) the equivalent Rabi frequency at resonance $\Lambda_R=\omega_P\sqrt{f_{\MF{L}}}/2$ becomes small compared to the bare cavity and TO phonon frequencies. In this regime of weakly coupled oscillators ($\Lambda_R/\omega_\mathrm{TO}\ll1$), the mixing fractions can be approximated by $h_l\approx|x_{\bm{q}}|^2$ and $h_p\approx|z_{\bm{q}}|^2$ as well as the polariton operator $\Pi_{\bm{q}}\approx x_{\bm{q}}a_{\bm{q}}+z_{\bm{q}}p_{\bm{q}}=\Pi_{\mathrm{l},\bm{q}}+\Pi_{\mathrm{p},\bm{q}}$. \cite{ciuti_quantum_2005}

\section{Polariton-ISB interaction}\label{sec:polisb}
The second step of our approach consists of describing the interaction between the phonon-polariton and the ISB transitions.
We express the full quantum Hamiltonian that describes the phonon-polariton-ISB system as follows
\begin{align}
 \hat{\Ham}_\mathrm{p-ISB}=&\sum_{\bm{q}}\hbar\omega_{\bm{q},\mathrm{p}}\Pi^\dagger_{{\bm{q}}}\Pi_{{\bm{q}}}+
 \sum_{j}\hbar\omega_{j}b_{j}^\dagger b_{j} \nonumber
 \\
 &+i\sum_{\bm{q},j}\hbar\Omega_{j,\q}\left(a_{-\q} - a_{\q}^\dagger \right)
 \left(b^\dagger_{j}+b_{j}\right) \nonumber\\
&+\sum_{\bm{q},j}\hbar\Xi_{j,\q}\left(p_{\q} + p_{-\q}^\dagger \right)\left(b^\dagger_{j}+b_{j}\right).
 \label{hamiltonian_polariton_ISB}
\end{align}
The first term in Eq.~(\ref{hamiltonian_polariton_ISB}) is the polaritonic part of the Hamiltonian with the approximate polariton operator $\Pi_{{\bm{q}}}$ already defined above, in either the upper or lower polariton branch. The second term contains the ISB part. Similarly to Eq.~\eqref{coupling_great_lambda}, the two last terms in (\ref{hamiltonian_polariton_ISB}) are the interaction components with respective coupling frequencies
\begin{align}
 \label{photon_coupling}
  \Omega_{j,\q}&=\frac{\omega_{P_j}}{2}\sqrt{\frac{\omega_{\text{opt,}\q}}{\omega_j}}
 \frac{\int f_{\bm{q}}(z)\xi_{j}(z)dz}{\sqrt{L_\mathrm{per}\int \xi_{j}^2(z)dz}},\\
 \Xi_{j,\q} &= \frac{\omega_{\MF{P_j}}\omega_{P}}{4\sqrt{\omega'\omega_j}}\frac{\int \xi_{\MF{L}}(z)\xi_j(z)dz}{\sqrt{\int \xi^2_j(z)dz\int \xi^2_{\MF{L}}(z)dz}}
 \label{phonon_coupling}
\end{align}
where $\omega_{P_j}$ is the ISB plasma frequency proportional to the injected carrier density \cite{todorov_intersubband_2012}.

Since the phonon-polariton mode and the ISB transitions are in the weak coupling regime, Fermi's golden rule can be applied to compute the emission rate, i.e., the gain cross section of the phonon-polariton-ISB system.
In a cavity containing $N_{\bm{q}}$ phonon-polaritons, we consider all the transitions $|ul,N_{\bm{q}}\rangle\rightarrow|n,N_{\bm{q}}+1\rangle$ \MF{with an electron initally in the upper laser state ($ul$) and finally in a lower energy level $n$,} that lead to the emission of a phonon-polariton and we calculate the total emission rate.
By Fermi's golden rule, then the emission rate becomes%
\begin{widetext}
\begin{equation}
 g(\omega_{\bm{q}})=\frac{2\pi}{\hbar}\sum_j\Big{|} \langle n,N_{\q}+1 | \\ \Big{[} i \hbar\Omega_{j,\q}\left( a_{-\q} - a_{\q}^\dagger \right)\left(b^\dagger_{j}-b_{j}\right)
 +\hbar\Xi_{j,\q}\left( p_{\q} + p_{-\q}^\dagger  \right)\left(b^\dagger_{j}-b_{j}\right)\Big{]}\\|ul,N_{\q}\rangle \Big{|}^2\delta(\omega-\omega_j).
 \label{fermi_golden_rule}
\end{equation}%
\end{widetext}
\noindent
Retaining only the terms in (\ref{fermi_golden_rule}) that describe an emission process (the ones that are proportional to  $a_{\q}^\dagger b_{j}$ and $p_{\q}^\dagger b_{j}$), we find the expression of the  total gain cross section
\begin{eqnarray}
  &g(\omega_{\bm{q}})=2\pi\hbar (N_{\bm{q}}+1)\sum_j\left|z^\ast_{\bm{q}}\Xi_{j,\q}-ix^\ast_{\bm{q}}\Omega_{j,\q}\right|^2\delta(\omega_j-\omega_{\bm{q}}) \nonumber \\
 & =2\pi\hbar (N_{\bm{q}}+1)\sum_j\big{|} |z_{\bm{q}}|\Xi_{j,\q}+|x_{\bm{q}}|\Omega_{j,\q}\big{|}^2\delta(\omega_j-\omega_{\bm{q}}) \nonumber
 \label{emission_rate}
\end{eqnarray}
as shown in Appendix \ref{sec:polstates}. Here special care needs to be taken to the phase between $x_{\q}$ and $z_{\q}$, since this is evidently crucial to the role of the mixed terms in Eq.~\eqref{emission_rate}. 
If we write $x_{\q} = |x_{\q}| e^{i\varphi}$, then $z_{\q} = |z_{\q}| e^{i(\varphi + \pi/2)}$ leading to the second line of Eq.~\eqref{emission_rate}. Thus, the mixed terms contribute constructively to the emission rate, if the couplings $\Omega_{j,\q}$ and $\Xi_{j,\q}$ have the same sign.
The $\delta$ function can be replaced by a Lorentzian function of characteristic width $\gamma$: $\delta(\omega_{j}-\omega_{\bm{q}}) \rightarrow \frac{\gamma/\pi}{(\omega_j-\omega_{\bm{q}})^2+\gamma^2}$.

Along the same lines, the optical losses of the device can be estimated from the respective photon ($\tau_\mathrm{cav}$) and phonon ($\tau_\mathrm{p}$) lifetimes in the cavity. Accounting again for the mixed nature of the polariton, the loss rate is written as
\begin{equation}
 \alpha(\omega_{\bm{q}})=\frac{h_l(\bm{q})}{\tau_\mathrm{cav}}+\frac{h_p(\bm{q})}{\tau_\mathrm{p}}.\label{eq:loss}
\end{equation}
where, from experimental studies values of phonon lifetimes, $\tau_p$ of 3.5 ps at 300 K and 7.8 ps at 77 K were determined \cite{bhatt_theoretical_1994} and $\tau_\mathrm{cav}$ can be readily estimated from the cavity losses. 

In the following section we will employ the developed theory to compute the phonon-polariton dispersion and gain in experimentally realizable 2D systems.

\section{Computational examples} \label{sec:examples}
\subsection{InGaAs-based resonant tunnelling diode} \label{sec:RTD}

\begin{figure}
 \includegraphics[width=\linewidth]{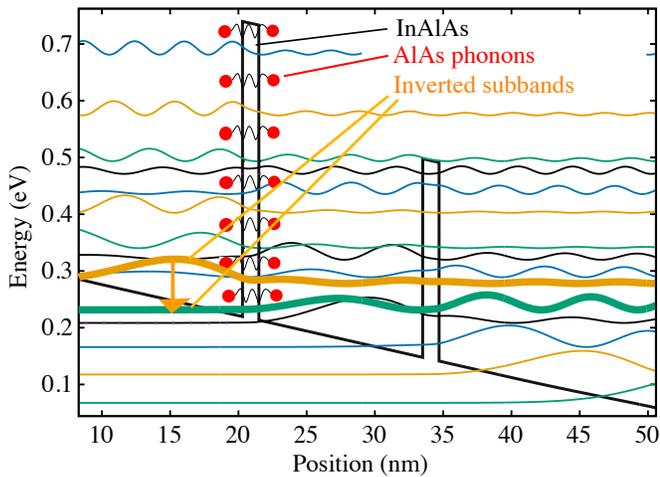}
 \caption{(Color online) Transport scheme of a phonon-polariton resonant tunnelling diode (RTD). The red oscillators represent the confining barrier of the TO phonon modes which at the same time serves as the tunnelling barrier. The electronic wavefunctions of the upper and the lower lasing states are plotted thick yellow and green lines, respectively, and overlap the TO phonon modes such that an additional gain arising from the TO phonon-ISB transition coupling is expected. Under the lasing bias, the energy spacing between the two lasing states is resonant with the TO phonon energy. The layer sequence of the structure in \AA~is 
\underline{250}/100/\textit{12}/120/\textbf{12}/50/\underline{5000}, where bold face denotes AlGaSb barriers, italic face denotes the AlInAs barrier, and the underlined layers are doped to $2\cdot10^{16}$ cm${}^{-3}$.}
\label{fig:RTD}
\end{figure}

As a first example, we consider a resonant tunnelling diode (RTD) structure. The benefit of such a structure for emission in the THz region, is that it can be heavily doped and thus have a large inversion, in addition to easily tuneable emission frequency by changing the layer widths. In addition, the simple layer structure provides an excellent starting point for a theoretical analysis of polariton gain in heterostructure. However, such devices typically have population inversion in regions of negative differential conductance (NDC), and can thus not operate in a serial configuration. In addition,  in the structure shown in Fig.~\ref{fig:RTD}, a four mono-layer thick InAlAs barrier serves as both the injection barrier of the RTD, giving rise to inversion between the level indicated by a thick yellow line and the two semi-bound states of the subsequent quantum well, as well as the confining layer for the AlAs phonons. In the following computations, we treat the four mono-layers as one effective layer with $\sigma=0.48$ nm and $\omega_P=17$ meV. The computed optical loss for this structure is $\sim 410$ cm\inv with a Au/Au double-metal waveguide. In comparison, we calculate a maximum optical gain of $\sim 800$ cm\inv, using a non-equilibrium Green's function model.\cite{wacker_nonequilibrium_2013}

\begin{figure}
\includegraphics[width = \linewidth]{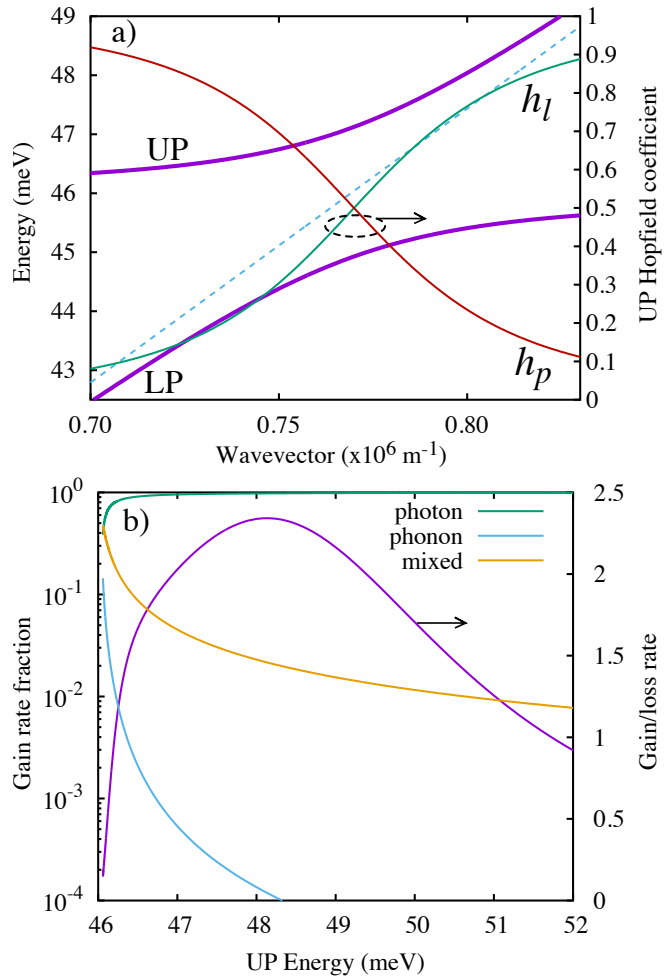}
\caption{(Color online) a) Calculated dispersion of light $\omega_{\bm{q}}\pm$ for the phonon-polariton RTD in Fig.~\ref{fig:RTD}. The polaritonic phonon ($h_p$) and photon ($h_l$) mixing fractions of as functions of the energy in the upper branch, are also shown with thin lines. The dashed line shows the bare cavity mode with 
$\omega =\frac{ck}{\sqrt{\epsilon_r}}$.
b) Gain fraction of the different gain components arising from the photon (green), phonon (blue) and mixed terms (orange) in (\ref{emission_rate}). The right axis shows the ratio of the gain to the losses in the UP branch.}
\label{fig:RTDdisp}
\end{figure}

From Eq.~\eqref{eigenvalue0}, we compute the phonon-polariton dispersion which is shown in Fig.~\ref{fig:RTDdisp} a). Due to the small filling factor $f_{\MF{L}}=1.67\cdot10^{-2}$, this structure exhibits a much smaller polaritonic gap than the one of bulk AlAs. 
%
%

Fig.~\ref{fig:RTDdisp} b) shows the contributions to the gain rate of Eq.~\eqref{emission_rate} from the photon, phonon, and mixed parts, as functions of the energy in the upper polariton branch.  For low energies, close to the polariton gap, the phonon fraction is maximal and decreases rapidly with increasing $\omega_{\bm{q}}$. Reversely, the photonic gain vanishes when $\omega\rightarrow\omega'$, but dominates at high frequencies as the photon fraction increases. Due to the small filling factor, the coupling ratio $\Xi_{\bm{q}}/\Omega_{\bm{q}}\ll1$ and the total gain is mostly dominated by the photonic gain. However, a maximum non-photonic gain of 20\% is already achieved at the frequency where gain overcomes the losses, despite a phonon extension of only a few monolayers.
Figure \ref{fig:RTDdisp} b) also shows the ratio $g/\alpha$ as a function of the energy in the UP branch. For the bias considered here, the bare optical gain is peaked at a frequency of $\hbar\omega=48.05$ meV. The loss rate being energy-dependent through the mixing fractions, dividing the gain by the losses shifts its maximum by 0.1 meV, which corresponds to the lasing energy of the device. While the phonon gain fraction at this energy is only $1.2\cdot10^{-4}$, the non-photonic contribution to the gain is still 2\% of the total gain. In addition, this structure has relatively low optical losses, and the contribution of the phonon part of the polariton is expected to be more important for structures where the optical losses are higher.

\subsection{InGaAs-based quantum cascade laser} \label{sec:QCL}

Our second example is a quantum cascade laser\cite{faist_quantum_1994} (QCL) where the TO phonons are provided by a barrier close to the inverted ISB transition. In contrast to RTDs, QCLs are reliable sources of coherent radiation in the THz frequency region, with a well proven growth and fabrication technique. In addition, operating at a bias of positive differential resistance, one QCL period can be repeated hundreds of times in a several $\mu$m thick structure, potentially allowing significantly more optical power to be extracted than from a single period RTD structure.

\begin{figure}
 \includegraphics[width=\linewidth]{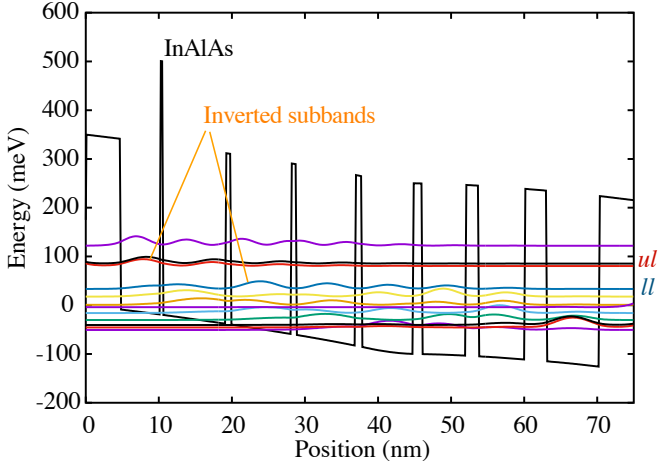}
 \caption{(Color online) Band structure and eigenstates of the proposed phonon-polariton QCL for an applied electric field of 18 kV/cm. The marked AlInAs barrier hosts the polaritons and overlaps the gain transition, from the upper laser state ($ul$) to the lower laser state ($ll$). The electrons are depopulated from the $ll$ into the 
 $ul$ of the next period via cascading down the potential wells through coherent tunnelling, as well as incoherent transport. The layer sequence in \AA~is, starting from the rightmost barrier, \textbf{48}/54/\textit{3}/86/\textbf{7.5}/82/\textbf{7.5}/81/\textbf{8.5}/71/\textbf{11.2}/61/\textbf{16}/\underline{64}/\textbf{30}/72, where bold face denotes GaAsSb barriers, italic font denotes the InAlAs barrier, and the underlined well is doped to $4.1\cdot10^{17}$ cm${}^{-3}$.}
 \label{fig:QCL}
 \end{figure}

In the structure in Fig.~\ref{fig:QCL}, a monolayer-thick AlInAs barrier plays the role of the phonon barrier in a InGaAs/GaAsSb active region\cite{ohtani_quantum_2016}. This barrier is placed where the the inverted subbands $ul$ and $ll$ have significant overlap, thus emitting phonon-polaritons via stimulated emission. In this bound-to-continuum design, the carriers are extracted from $ll$ in a cascade ending on the black state of lowest energy in Fig.~\ref{fig:QCL}, where it is subsequently injected in to the $ul$ state of the next period of the QCL. For this structure, the electron transport is calculated in a density matrix approach.\cite{terazzi_density_2010}
 \begin{figure}
 \includegraphics[width=\linewidth]{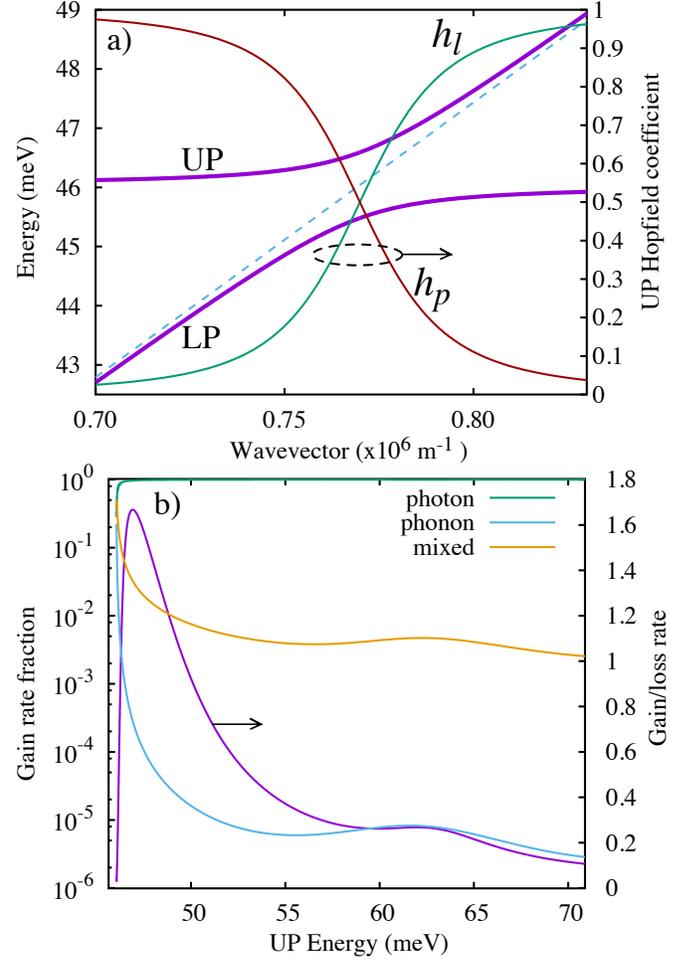}
 \caption{(Color online) a) Dispersion relation of the proposed phonon-polariton QCL, where the design frequency belongs to the upper branch, as well as phonon and phonon hop field coefficients (thin lines). The dashed line shows the bare cavity mode with $\omega = \frac{ck}{\sqrt{\epsilon_r}}$.
 b) Gain fraction of the different gain components arising from the photonic part (green), the phononic part (blue) and the mixed terms (orange) in (\ref{emission_rate}). The right axis shows the ratio of the gain to the losses of the UP polariton mode.}
 \label{fig:QCLdisp}
 \end{figure}
 
The calculated dispersion and mixing fractions are shown in Fig.~\ref{fig:QCLdisp} a). In this structure, $f_{\MF{L}}=4.3\cdot10^{-3}$ is even smaller than for the RTD. However the design frequency is adjusted to be close to the maximum splitting between the branches, where the fraction of the phonon to photon Hopfield coefficitent is close to 50\%. For this and slightly higher frequencies, the design has a phonon fraction of about $3\cdot10^{-4}$ of the total gain, while the total non-photonic gain accounts for $\sim 3$\%, as seen in Fig.~\ref{fig:QCLdisp} b). In this figure we also show the ratio of the calculated gain to the losses from Eq.~\eqref{eq:loss}, and we find the maximum value at an energy slightly blue shifted from the design frequency. The second, lower, peak at 62 meV, arises due to emission to a lower electronic state which has less overlap with the upper laser level.

Despite the fact that the gain of these devices remains mainly dominated by the standard dipole coupling, there is room for increasing the phononic contribution. An optimized design with a suitable location and thickness of the phonon layer could lead to larger overlaps between the phonon and ISB microcurrents.  The choice of a material with a larger polariton gap (with larger $\omega_P$), such as ZnO/ZnMgO or GaN/AlN, could dramatically increase the phonon part of the gain. In particular for structures with large optical losses compared to optical gain, the phonon contribution to the polaritonic gain can then increase the stimulated emission rate and thereby the conversion efficiency of electrical power into power radiated in the electric field.
 
\section{Conclusion}

In conclusion, we have developed a quantum approach for the description of gain in phonon-polariton lasers. Compared to an effective dielectric model, \cite{henry_raman_1965} this formalism is more appropriate for the description of confined modes in thin layers and has the advantage to provide a complete physical insight of the system, especially by directly giving the phonon and photon fractions of the lasing modes that are the key parameters for the gain computation. Our model can be applied to a wide variety of designs and material systems, offering a wide set of possibilities for the optimization of future phonon-polariton QCLs. As a demonstration of the flexibility of the model, we have proposed and simulated resonant tunnelling diodes and quantum cascade lasers made from the conventional InGaAs/InAlAs/InGaSb material system, as well as the less explored ZnO material system. While the former two structures show a small non-photonic contribution to the gain of $\sim$ 10 \%, this number can be increased by employing more phonon material to increase the filling factor, or using other material systems with larger phonon plasma frequency, such as ZnO/ZnMgO or GaN/AlN. 

\acknowledgments{This work is partly supported by the ERC project MUSiC as well as by the NCCR QSIT. JF thank A. Vasanelli, S. De Liberato and J. B. Khurgin for very fruitful discussions.
}

\appendix

\section{Phonon polarization in second quantization} \label{sec:phonapp}

\MF{For modelling the lattice vibrations in thin layers at $z = z_i$, we expand the 
collective lattice vibrations in the basis of vibrational harmonic oscillator modes $\Psi_{\alpha,ij}(R) = \psi_{\alpha,ij}(z)\chi_{\alpha,ij}(\rr)$, where $j$ labels the in-plane coordinate and $\alpha$ is the harmonic oscillator excitation:
\begin{eqnarray}
\hat{\Psi}^\dagger &=& \frac{1}{\sqrt{S}}\sum_{\alpha,ij} d^\dagger_{\alpha,ij} \psi_{\alpha,i,j}(z)\chi_{\alpha,i,j}(\rr),
\end{eqnarray}
where $S$ is the sample area and $d^\dagger_{\alpha,ij}$ the creation operator for a lattice vibration excitation. Due to the rotational symmetrty in the $x$-$y$ plane, we write the total wave function $\Psi(R)$ as a product between the harmonic oscillator wave function $\psi_\alpha(z)$, and the in-plane (periodic) wave function $\chi_{\alpha,j}(\rr)$.}
As in Ref.~\onlinecite{todorov_intersubband_2012}, the polarization is found via its relation to the current density operator
\begin{equation}
\hat{J}_z(R) = \frac{1}{i\hbar}[\hat{P}_{L,z},{\cal H}], \label{eq:polheisenberg}
\end{equation}
defined by
\MF{
\begin{eqnarray}
\hat{J}_z(R) &=&\frac{i\hbar e}{2M}\left(\hat{\Psi}^\dagger \frac{\partial }{\partial z} \hat{\Psi} - (\frac{\partial }{\partial z} \hat{\Psi}^\dagger )\hat{\Psi}\right) \\
 &=& \frac{i\hbar e}{2SM}\sum_{\alpha\beta}\sum_{ii',jj'} \xi_{\alpha\beta}^{ii'}(z) \chi_{\alpha j}^\ast\chi_{\beta j'} d_{\alpha,ij}^{\dagger}d_{\beta,i'j'},\nonumber
\end{eqnarray}
}
where $M$ is parametrising the inertia of the ions. Here, the phonon micro current is defined as
\begin{equation}
\xi^{ii'}_{\alpha\beta}(z) = \phi_{\alpha,i}(z)\frac{\partial}{\partial z}\phi^{\ast}_{\beta,i'}(z)-\phi^{\ast}_{\beta,i'}(z)\frac{\partial}{\partial z}\phi_{\alpha,i}(z).
\end{equation}
The only allowed transitions of the harmonic oscillators are those with $|\alpha-\beta|=1$, and we will consider only the lowest excitation with $(\alpha,\beta)\in \{0,1\}$. Then, the current density operator  becomes
\MF{
\begin{eqnarray}
\hat{J}_z(R) &=& \frac{i\hbar e}{2SM} \sum_{i,jj'} \xi_{10}^{ii}(z)\times \\\nonumber
&\times& (\chi_{1,ij}^\ast\chi_{0,ij'}d_{1,ij}^\dagger d_{0,ij'}- \chi_{1,ij'}\chi_{0,ij}^\ast d_{0,ij} d_{1,ij'}^\dagger).
\end{eqnarray}
}%
\MF{Here, we neglected the mixing of different layers $i\neq i'$, which will give a very small contribution if the layers are separated by a few standard deviations $\sigma$.}
The terms in the bracket are periodic in the plane with period \MF{$a_{x/y} = \frac{2\pi}{q_{x/y}}$}, since \MF{we are interested in solutions where} the polarization is a travelling wave with momentum $\q$, and the first term is Fourier expanded to (the second term is just the complex conjugate of the first one)
\MF{
\begin{equation}
\sum_{jj'} \chi_{1,ij}^\ast\chi_{0,ij'}d_{1,ij}^\dagger d_{0,ij'} \equiv \sum_{\q} d^\dagger_{i,\q} e^{-i\q\rr}.\label{eq:qidentity}
\end{equation}
Thus, the current density becomes
\begin{equation}
\hat{J}_z(R) = \frac{i\hbar e}{2SM} \sum_{i,\q} \xi_{L}(z) e^{i\q \rr}(d_{i,-\q}^\dagger - d_{i,\q}),
\end{equation}
where $\xi_{L}(z) \equiv \xi_{10}^{ii}(z)$}.
Using the commutation relations \MF{$[d_{i\q}^\dagger,{\cal \hat{H}}] = -\hbar\omega_\text{TO} d_{i\q}^\dagger$ and $[d_{i\q},{\cal \hat{H}}] = +\hbar\omega_\text{TO} d_{i\q}$} together with Eq.~\eqref{eq:polheisenberg}, we find the polarization density operator of Eq.~\eqref{eq:phonpol}.

The similar form of the phonon polarization to the electronic one, prompts us to write the Hamiltonian for the TO phonon as
\begin{equation}
{\cal H}_L = \sum_{i\q} \hbar\omega_\text{TO} d^\dagger_{i,\q} d_{i,\q}. \label{eq:Hq}
\end{equation}
Inserting Eq.~\eqref{eq:phonpol} in place of the classical polarization in the classical Hamiltonian (the last line of Eq.~\eqref{eq:hclass}) gives
\begin{eqnarray}
{\cal H}_L &=& \int d^3 r \frac{1}{2\chi_L}(\omega_\text{TO}^2\P_L^2 + \dot{\P}_L^2) \\ 
&=& \frac{1}{2\chi_L}\frac{\hbar^2e^2}{SM^2}\int \xi^2_{\MF{L}}(z) dz \sum_{i\q} d_{i,\q}^\dagger d_{i,\q}
\end{eqnarray}
(apart from the constant vacuum energy shift which is irrelevant here). This is equal to Eq.~\eqref{eq:Hq} if we identify the phonon plasma frequency $\omega_P^2=\chi_L/\epsilon_0\epsilon_r$ as in Eq.~\eqref{eq:plasmaf}.

\section{Phonon-polariton eigenstates}\label{sec:polstates}

In order to compute the polariton scattering rates, we need to express the polariton states in terms it's phonon and photon constituents.
The polariton states can be calculated by repeated application of the polariton creation operator $\Pi_{\q}^\dagger$ on the vacuum state $|0\rangle$ as 
\begin{equation}
|N_{\q}\rangle = C_{N_{\q}}(\Pi^\dagger_{\q})^{N_{\q}}|0\rangle, \label{eq:Nq}
\end{equation}
with a normalization constant $C_{N_{\q}}$, to be determined. From now on we suppress the index ${\q}$ for ease of notation. Eq.~\eqref{eq:Nq} can be rewritten by noting that $(\Pi^\dagger)^N = (x^\ast a^\dagger + z^\ast p^\dagger)^N$ and using the binomial formula as
\begin{eqnarray}
|N\rangle &=& C_N \sum_{k=0}^N \frac{N!}{k!(N-k)!}(x^\ast)^k (z^\ast)^{N-k} (a^\dagger)^k (p^\dagger)^{N-k} |0\rangle \nonumber\\
&=& C_N \sum_{k=0}^N \frac{N!}{\sqrt{k!(N-k)!}}(x^\ast)^k (z^\ast)^{N-k} |k,N-k\rangle
\end{eqnarray}
where $|n,m\rangle \equiv |n\rangle_\text{phot}\otimes|m\rangle_\text{phon}$ span the Hilbert space of the phonon-polariton. Here, we used the approximation of small coupling strength $\Lambda_R/\omega_\mathrm{TO}\ll1$. However, the resulting $|N\rangle$ will be the same also without this approximation, since terms like $a_{-\q}a^\dagger_{\q}|0\rangle=0$. The normalization constant is found by solving
\begin{eqnarray}
&1 = \langle N | N \rangle = \\ &|C_N|^2\sum_{k,k'}^N \frac{ N!N! x^{k'}(x^\ast)^{k} z^{N-k'}(z^\ast)^{N-k} }{\sqrt{k!(N-k)!}\sqrt{k'!(N-k')!}} \langle k', N-k' | k, N-k \rangle.\nonumber
\end{eqnarray}
The bra-ket gives $\delta_{kk'}$, and again using the binomial formula, we find
\begin{equation}
|C_N|^2 \sum_k \frac{N!N!}{k!(N-k)!}|x|^{2k}|z|^{2N-2k} = |C_N|^2 N! (|x|^2 + |z|^2)^N.
\end{equation}
By definition, $|x|^2+|z|^2 = 1$, and we readily find that $C_N = 1/\sqrt{N!}$, and
\begin{equation}
|N\rangle = \sum_{k=0}^N \sqrt{ \frac{ N! }{ k!(N-k)! } } (x^\ast)^k (z^\ast)^{N-k} |k,N-k\rangle.
\end{equation}
We can easily check that the number operator gives the correct result by using $[a,(a^\dagger)^N] = N(a^\dagger)^{N-1}$:
\begin{eqnarray}
\Pi^\dagger\Pi|N\rangle &=& \Pi^\dagger\Pi (\Pi^\dagger)^N C_N |0\rangle \nonumber \\ &=& \Pi^\dagger N(\Pi^\dagger)^{N-1} C_N|0\rangle = N|N\rangle.
\end{eqnarray}

Now lets calculate the emission rate $\Gamma^\text{em}$. For this, we need to compute terms with
\begin{eqnarray}
\langle N+1 | a^\dagger |N\rangle &=& \sum_{k=0}^{N} \frac{ (x^\ast)^{2k+1} (z^\ast)^{2N-2k} } { k!(n-k)! }\sqrt{N!(N+1)!} \nonumber\\ &\equiv& x^\ast\sqrt{N+1}\sum_{k=0}^N C_{k,N}^2
\end{eqnarray}
where we defined $|N\rangle \equiv \sum_k C_{k,N} |k, N-k\rangle$. Similarly, we find that
\begin{equation}
\langle N+1 | p^\dagger |N\rangle = z^\ast\sqrt{N+1}\sum_{k=0}^{N} C_{k,N}^2.
\end{equation}
The normalization of $|N\rangle$ means that $\sum_{k=0}^N C^2_{k,N} = 1$ and so
\begin{eqnarray}
\langle N+1 | a^\dagger |N\rangle = \frac{\langle N+1 | \Pi_{l}^\dagger |N\rangle}{x^\ast}  &=& x^\ast\sqrt{N+1}\label{eq:phot_element} \\
\langle N+1 | p^\dagger |N\rangle = \frac{\langle N+1 | \Pi_{p}^\dagger |N\rangle}{z^\ast}  &=& z^\ast\sqrt{N+1}\label{eq:phon_element} 
\end{eqnarray}
Eqs.~(\ref{eq:phot_element}-\ref{eq:phon_element}) inserted in Eq.~\eqref{fermi_golden_rule} give the emission rate in Eq.~\eqref{emission_rate}.


\begin{thebibliography}{32}%
\makeatletter
\providecommand \@ifxundefined [1]{%
 \@ifx{#1\undefined}
}%
\providecommand \@ifnum [1]{%
 \ifnum #1\expandafter \@firstoftwo
 \else \expandafter \@secondoftwo
 \fi
}%
\providecommand \@ifx [1]{%
 \ifx #1\expandafter \@firstoftwo
 \else \expandafter \@secondoftwo
 \fi
}%
\providecommand \natexlab [1]{#1}%
\providecommand \enquote  [1]{``#1''}%
\providecommand \bibnamefont  [1]{#1}%
\providecommand \bibfnamefont [1]{#1}%
\providecommand \citenamefont [1]{#1}%
\providecommand \href@noop [0]{\@secondoftwo}%
\providecommand \href [0]{\begingroup \@sanitize@url \@href}%
\providecommand \@href[1]{\@@startlink{#1}\@@href}%
\providecommand \@@href[1]{\endgroup#1\@@endlink}%
\providecommand \@sanitize@url [0]{\catcode `\\12\catcode `\$12\catcode
  `\&12\catcode `\#12\catcode `\^12\catcode `\_12\catcode `\%12\relax}%
\providecommand \@@startlink[1]{}%
\providecommand \@@endlink[0]{}%
\providecommand \url  [0]{\begingroup\@sanitize@url \@url }%
\providecommand \@url [1]{\endgroup\@href {#1}{\urlprefix }}%
\providecommand \urlprefix  [0]{URL }%
\providecommand \Eprint [0]{\href }%
\providecommand \doibase [0]{http://dx.doi.org/}%
\providecommand \selectlanguage [0]{\@gobble}%
\providecommand \bibinfo  [0]{\@secondoftwo}%
\providecommand \bibfield  [0]{\@secondoftwo}%
\providecommand \translation [1]{[#1]}%
\providecommand \BibitemOpen [0]{}%
\providecommand \bibitemStop [0]{}%
\providecommand \bibitemNoStop [0]{.\EOS\space}%
\providecommand \EOS [0]{\spacefactor3000\relax}%
\providecommand \BibitemShut  [1]{\csname bibitem#1\endcsname}%
\let\auto@bib@innerbib\@empty
\bibitem [{\citenamefont {Hopfield}(1958)}]{hopfield_theory_1958}%
  \BibitemOpen
  \bibfield  {author} {\bibinfo {author} {\bibfnamefont {J.~J.}\ \bibnamefont
  {Hopfield}},\ }\href {\doibase 10.1103/PhysRev.112.1555} {\bibfield
  {journal} {\bibinfo  {journal} {Phys. Rev.}\ }\textbf {\bibinfo {volume}
  {112}},\ \bibinfo {pages} {1555} (\bibinfo {year} {1958})}\BibitemShut
  {NoStop}%
\bibitem [{\citenamefont {Carusotto}\ and\ \citenamefont
  {Ciuti}(2013)}]{carusotto_quantum_2013}%
  \BibitemOpen
  \bibfield  {author} {\bibinfo {author} {\bibfnamefont {I.}~\bibnamefont
  {Carusotto}}\ and\ \bibinfo {author} {\bibfnamefont {C.}~\bibnamefont
  {Ciuti}},\ }\href {\doibase 10.1103/RevModPhys.85.299} {\bibfield  {journal}
  {\bibinfo  {journal} {Rev. Mod. Phys.}\ }\textbf {\bibinfo {volume} {85}},\
  \bibinfo {pages} {299} (\bibinfo {year} {2013})}\BibitemShut {NoStop}%
\bibitem [{\citenamefont {Deng}\ \emph {et~al.}(2002)\citenamefont {Deng},
  \citenamefont {Weihs}, \citenamefont {Santori}, \citenamefont {Bloch},\ and\
  \citenamefont {Yamamoto}}]{deng_condensation_2002}%
  \BibitemOpen
  \bibfield  {author} {\bibinfo {author} {\bibfnamefont {H.}~\bibnamefont
  {Deng}}, \bibinfo {author} {\bibfnamefont {G.}~\bibnamefont {Weihs}},
  \bibinfo {author} {\bibfnamefont {C.}~\bibnamefont {Santori}}, \bibinfo
  {author} {\bibfnamefont {J.}~\bibnamefont {Bloch}}, \ and\ \bibinfo {author}
  {\bibfnamefont {Y.}~\bibnamefont {Yamamoto}},\ }\href {\doibase
  10.1126/science.1074464} {\bibfield  {journal} {\bibinfo  {journal}
  {Science}\ }\textbf {\bibinfo {volume} {298}},\ \bibinfo {pages} {199}
  (\bibinfo {year} {2002})}\BibitemShut {NoStop}%
\bibitem [{\citenamefont {Kasprzak}\ \emph {et~al.}(2006)\citenamefont
  {Kasprzak}, \citenamefont {Richard}, \citenamefont {Kundermann},
  \citenamefont {Baas}, \citenamefont {Jeambrun}, \citenamefont {Keeling},
  \citenamefont {Marchetti}, \citenamefont {Szyma{\'n}ska}, \citenamefont
  {Andr{\'e}}, \citenamefont {Staehli}, \citenamefont {Savona}, \citenamefont
  {Littlewood}, \citenamefont {Deveaud},\ and\ \citenamefont
  {Dang}}]{kasprzak_boseeinstein_2006}%
  \BibitemOpen
  \bibfield  {author} {\bibinfo {author} {\bibfnamefont {J.}~\bibnamefont
  {Kasprzak}}, \bibinfo {author} {\bibfnamefont {M.}~\bibnamefont {Richard}},
  \bibinfo {author} {\bibfnamefont {S.}~\bibnamefont {Kundermann}}, \bibinfo
  {author} {\bibfnamefont {A.}~\bibnamefont {Baas}}, \bibinfo {author}
  {\bibfnamefont {P.}~\bibnamefont {Jeambrun}}, \bibinfo {author}
  {\bibfnamefont {J.~M.~J.}\ \bibnamefont {Keeling}}, \bibinfo {author}
  {\bibfnamefont {F.~M.}\ \bibnamefont {Marchetti}}, \bibinfo {author}
  {\bibfnamefont {M.~H.}\ \bibnamefont {Szyma{\'n}ska}}, \bibinfo {author}
  {\bibfnamefont {R.}~\bibnamefont {Andr{\'e}}}, \bibinfo {author}
  {\bibfnamefont {J.~L.}\ \bibnamefont {Staehli}}, \bibinfo {author}
  {\bibfnamefont {V.}~\bibnamefont {Savona}}, \bibinfo {author} {\bibfnamefont
  {P.~B.}\ \bibnamefont {Littlewood}}, \bibinfo {author} {\bibfnamefont
  {B.}~\bibnamefont {Deveaud}}, \ and\ \bibinfo {author} {\bibfnamefont
  {L.~S.}\ \bibnamefont {Dang}},\ }\href {\doibase 10.1038/nature05131}
  {\bibfield  {journal} {\bibinfo  {journal} {Nature}\ }\textbf {\bibinfo
  {volume} {443}},\ \bibinfo {pages} {409} (\bibinfo {year}
  {2006})}\BibitemShut {NoStop}%
\bibitem [{\citenamefont {Deng}\ \emph {et~al.}(2010)\citenamefont {Deng},
  \citenamefont {Haug},\ and\ \citenamefont
  {Yamamoto}}]{deng_exciton-polariton_2010}%
  \BibitemOpen
  \bibfield  {author} {\bibinfo {author} {\bibfnamefont {H.}~\bibnamefont
  {Deng}}, \bibinfo {author} {\bibfnamefont {H.}~\bibnamefont {Haug}}, \ and\
  \bibinfo {author} {\bibfnamefont {Y.}~\bibnamefont {Yamamoto}},\ }\href
  {\doibase 10.1103/RevModPhys.82.1489} {\bibfield  {journal} {\bibinfo
  {journal} {Rev. Mod. Phys.}\ }\textbf {\bibinfo {volume} {82}},\ \bibinfo
  {pages} {1489} (\bibinfo {year} {2010})}\BibitemShut {NoStop}%
\bibitem [{\citenamefont {Berini}\ and\ \citenamefont
  {De~Leon}(2012)}]{berini_surface_2012}%
  \BibitemOpen
  \bibfield  {author} {\bibinfo {author} {\bibfnamefont {P.}~\bibnamefont
  {Berini}}\ and\ \bibinfo {author} {\bibfnamefont {I.}~\bibnamefont
  {De~Leon}},\ }\href {\doibase 10.1038/nphoton.2011.285} {\bibfield  {journal}
  {\bibinfo  {journal} {Nat. Photonics}\ }\textbf {\bibinfo {volume} {6}},\
  \bibinfo {pages} {16} (\bibinfo {year} {2012})}\BibitemShut {NoStop}%
\bibitem [{\citenamefont {Colombelli}\ \emph {et~al.}(2005)\citenamefont
  {Colombelli}, \citenamefont {Ciuti}, \citenamefont {Chassagneux},\ and\
  \citenamefont {Sirtori}}]{colombelli_quantum_2005}%
  \BibitemOpen
  \bibfield  {author} {\bibinfo {author} {\bibfnamefont {R.}~\bibnamefont
  {Colombelli}}, \bibinfo {author} {\bibfnamefont {C.}~\bibnamefont {Ciuti}},
  \bibinfo {author} {\bibfnamefont {Y.}~\bibnamefont {Chassagneux}}, \ and\
  \bibinfo {author} {\bibfnamefont {C.}~\bibnamefont {Sirtori}},\ }\href
  {\doibase 10.1088/0268-1242/20/10/001} {\bibfield  {journal} {\bibinfo
  {journal} {Semicond. Sci. Technol.}\ }\textbf {\bibinfo {volume} {20}},\
  \bibinfo {pages} {985} (\bibinfo {year} {2005})}\BibitemShut {NoStop}%
\bibitem [{\citenamefont {Huber}\ \emph {et~al.}(2005)\citenamefont {Huber},
  \citenamefont {Ocelic}, \citenamefont {Kazantsev},\ and\ \citenamefont
  {Hillenbrand}}]{huber_near-field_2005}%
  \BibitemOpen
  \bibfield  {author} {\bibinfo {author} {\bibfnamefont {A.}~\bibnamefont
  {Huber}}, \bibinfo {author} {\bibfnamefont {N.}~\bibnamefont {Ocelic}},
  \bibinfo {author} {\bibfnamefont {D.}~\bibnamefont {Kazantsev}}, \ and\
  \bibinfo {author} {\bibfnamefont {R.}~\bibnamefont {Hillenbrand}},\ }\href
  {\doibase 10.1063/1.2032595} {\bibfield  {journal} {\bibinfo  {journal}
  {Appl. Phys. Lett.}\ }\textbf {\bibinfo {volume} {87}},\ \bibinfo {pages}
  {081103} (\bibinfo {year} {2005})}\BibitemShut {NoStop}%
\bibitem [{\citenamefont {Korobkin}\ \emph {et~al.}(2007)\citenamefont
  {Korobkin}, \citenamefont {Urzhumov}, \citenamefont {Neuner}, \citenamefont
  {Zorman}, \citenamefont {Zhang}, \citenamefont {Mayergoyz},\ and\
  \citenamefont {Shvets}}]{korobkin_mid-infrared_2007}%
  \BibitemOpen
  \bibfield  {author} {\bibinfo {author} {\bibfnamefont {D.}~\bibnamefont
  {Korobkin}}, \bibinfo {author} {\bibfnamefont {Y.~A.}\ \bibnamefont
  {Urzhumov}}, \bibinfo {author} {\bibfnamefont {B.}~\bibnamefont {Neuner}},
  \bibinfo {author} {\bibfnamefont {C.}~\bibnamefont {Zorman}}, \bibinfo
  {author} {\bibfnamefont {Z.}~\bibnamefont {Zhang}}, \bibinfo {author}
  {\bibfnamefont {I.~D.}\ \bibnamefont {Mayergoyz}}, \ and\ \bibinfo {author}
  {\bibfnamefont {G.}~\bibnamefont {Shvets}},\ }\href {\doibase
  10.1007/s00339-007-4084-8} {\bibfield  {journal} {\bibinfo  {journal} {Appl.
  Phys. A-Mater. Sci. Process.}\ }\textbf {\bibinfo {volume} {88}},\ \bibinfo
  {pages} {605} (\bibinfo {year} {2007})}\BibitemShut {NoStop}%
\bibitem [{\citenamefont {De~Liberato}\ and\ \citenamefont
  {Ciuti}(2012)}]{de_liberato_quantum_2012}%
  \BibitemOpen
  \bibfield  {author} {\bibinfo {author} {\bibfnamefont {S.}~\bibnamefont
  {De~Liberato}}\ and\ \bibinfo {author} {\bibfnamefont {C.}~\bibnamefont
  {Ciuti}},\ }\href {\doibase 10.1103/PhysRevB.85.125302} {\bibfield  {journal}
  {\bibinfo  {journal} {Phys. Rev. B}\ }\textbf {\bibinfo {volume} {85}},\
  \bibinfo {pages} {125302} (\bibinfo {year} {2012})}\BibitemShut {NoStop}%
\bibitem [{\citenamefont {Dini}\ \emph {et~al.}(2003)\citenamefont {Dini},
  \citenamefont {K{\"o}hler}, \citenamefont {Tredicucci}, \citenamefont
  {Biasiol},\ and\ \citenamefont {Sorba}}]{dini_microcavity_2003}%
  \BibitemOpen
  \bibfield  {author} {\bibinfo {author} {\bibfnamefont {D.}~\bibnamefont
  {Dini}}, \bibinfo {author} {\bibfnamefont {R.}~\bibnamefont {K{\"o}hler}},
  \bibinfo {author} {\bibfnamefont {A.}~\bibnamefont {Tredicucci}}, \bibinfo
  {author} {\bibfnamefont {G.}~\bibnamefont {Biasiol}}, \ and\ \bibinfo
  {author} {\bibfnamefont {L.}~\bibnamefont {Sorba}},\ }\href {\doibase
  10.1103/PhysRevLett.90.116401} {\bibfield  {journal} {\bibinfo  {journal}
  {Phys. Rev. Lett.}\ }\textbf {\bibinfo {volume} {90}},\ \bibinfo {pages}
  {116401} (\bibinfo {year} {2003})}\BibitemShut {NoStop}%
\bibitem [{\citenamefont {Ciuti}\ \emph {et~al.}(2005)\citenamefont {Ciuti},
  \citenamefont {Bastard},\ and\ \citenamefont
  {Carusotto}}]{ciuti_quantum_2005}%
  \BibitemOpen
  \bibfield  {author} {\bibinfo {author} {\bibfnamefont {C.}~\bibnamefont
  {Ciuti}}, \bibinfo {author} {\bibfnamefont {G.}~\bibnamefont {Bastard}}, \
  and\ \bibinfo {author} {\bibfnamefont {I.}~\bibnamefont {Carusotto}},\ }\href
  {\doibase 10.1103/PhysRevB.72.115303} {\bibfield  {journal} {\bibinfo
  {journal} {Phys. Rev. B}\ }\textbf {\bibinfo {volume} {72}},\ \bibinfo
  {pages} {115303} (\bibinfo {year} {2005})}\BibitemShut {NoStop}%
\bibitem [{\citenamefont {Power}\ and\ \citenamefont
  {Zienau}(1957)}]{power_radiative_1957}%
  \BibitemOpen
  \bibfield  {author} {\bibinfo {author} {\bibfnamefont {E.}~\bibnamefont
  {Power}}\ and\ \bibinfo {author} {\bibfnamefont {S.}~\bibnamefont {Zienau}},\
  }\href {\doibase 10.1007/BF02827754} {\bibfield  {journal} {\bibinfo
  {journal} {Nuovo Cimento}\ }\textbf {\bibinfo {volume} {6}},\ \bibinfo
  {pages} {7} (\bibinfo {year} {1957})}\BibitemShut {NoStop}%
\bibitem [{\citenamefont {Woolley}(1971)}]{woolley_molecular_1971}%
  \BibitemOpen
  \bibfield  {author} {\bibinfo {author} {\bibfnamefont {R.}~\bibnamefont
  {Woolley}},\ }\href {\doibase 10.1098/rspa.1971.0049} {\bibfield  {journal}
  {\bibinfo  {journal} {Proceedings of the Royal Society of London Series
  a-Mathematical and Physical Sciences}\ }\textbf {\bibinfo {volume} {321}},\
  \bibinfo {pages} {557} (\bibinfo {year} {1971})}\BibitemShut {NoStop}%
\bibitem [{\citenamefont {Todorov}\ and\ \citenamefont
  {Sirtori}(2012)}]{todorov_intersubband_2012}%
  \BibitemOpen
  \bibfield  {author} {\bibinfo {author} {\bibfnamefont {Y.}~\bibnamefont
  {Todorov}}\ and\ \bibinfo {author} {\bibfnamefont {C.}~\bibnamefont
  {Sirtori}},\ }\href {\doibase 10.1103/PhysRevB.85.045304} {\bibfield
  {journal} {\bibinfo  {journal} {Phys. Rev. B}\ }\textbf {\bibinfo {volume}
  {85}},\ \bibinfo {pages} {045304} (\bibinfo {year} {2012})}\BibitemShut
  {NoStop}%
\bibitem [{\citenamefont {Hopfield}(1995)}]{hopfield_aspects_1995}%
  \BibitemOpen
  \bibfield  {author} {\bibinfo {author} {\bibfnamefont {J.~J.}\ \bibnamefont
  {Hopfield}},\ }in\ \href
  {https://link.springer.com/chapter/10.1007/978-1-4615-1963-8_32} {\emph
  {\bibinfo {booktitle} {Confined {Electrons} and {Photons}}}},\ \bibinfo
  {series and number} {{NATO} {ASI} {Series}}\ (\bibinfo  {publisher}
  {Springer, Boston, MA},\ \bibinfo {year} {1995})\ pp.\ \bibinfo {pages}
  {771--782} \BibitemShut
  {NoStop}%
\bibitem [{\citenamefont {Gubbin}\ \emph {et~al.}(2016)\citenamefont {Gubbin},
  \citenamefont {Maier},\ and\ \citenamefont
  {De~Liberato}}]{gubbin_real-space_2016}%
  \BibitemOpen
  \bibfield  {author} {\bibinfo {author} {\bibfnamefont {C.~R.}\ \bibnamefont
  {Gubbin}}, \bibinfo {author} {\bibfnamefont {S.~A.}\ \bibnamefont {Maier}}, \
  and\ \bibinfo {author} {\bibfnamefont {S.}~\bibnamefont {De~Liberato}},\
  }\href {\doibase 10.1103/PhysRevB.94.205301} {\bibfield  {journal} {\bibinfo
  {journal} {Phys. Rev. B}\ }\textbf {\bibinfo {volume} {94}},\ \bibinfo
  {pages} {205301} (\bibinfo {year} {2016})}\BibitemShut {NoStop}%
\bibitem [{\citenamefont {{Christopher R. Gubbin}}\ \emph
  {et~al.}(2017)\citenamefont {{Christopher R. Gubbin}}, \citenamefont {{Stefan
  A. Maier}},\ and\ \citenamefont {{Simone De
  Liberato}}}]{christopher_r._gubbin_theoretical_2017}%
  \BibitemOpen
  \bibfield  {author} {\bibinfo {author} {\bibnamefont {{C.~R.
  Gubbin}}}, \bibinfo {author} {\bibnamefont {{S.~A. Maier}}}, \ and\
  \bibinfo {author} {\bibnamefont {{S.~De Liberato}}},\ }
  \href{\doibase 10.1103/PhysRevB.95.035313} {\bibfield  {journal}
  {\bibinfo  {journal} {Physical Review B}\ }\textbf {\bibinfo {volume} {95}},\
  \bibinfo {pages} {035313} (\bibinfo {year} {2017})}\BibitemShut {NoStop}%
\bibitem [{\citenamefont {Dzedolik}(2016)}]{dzedolik_phonon-polaritons_2016}%
  \BibitemOpen
  \bibfield  {author} {\bibinfo {author} {\bibfnamefont {I.~V.}\ \bibnamefont
  {Dzedolik}},\ }in\ \href
  {https://link.springer.com/chapter/10.1007/978-94-017-7315-7_1} {\emph
  {\bibinfo {booktitle} {Contemporary {Optoelectronics}}}},\ \bibinfo {series
  and number} {Springer {Series} in {Optical} {Sciences}}\ (\bibinfo
  {publisher} {Springer, Dordrecht},\ \bibinfo {year} {2016})\ pp.\ \bibinfo
  {pages} {3--23}\BibitemShut {NoStop}%
\bibitem [{\citenamefont {G{\'o}mez-Urrea}\ \emph {et~al.}(2017)\citenamefont
  {G{\'o}mez-Urrea}, \citenamefont {Duque}, \citenamefont
  {P{\'e}rez-Quintana},\ and\ \citenamefont
  {Mora-Ramos}}]{gomez-urrea_light_2017}%
  \BibitemOpen
  \bibfield  {author} {\bibinfo {author} {\bibfnamefont {H.~A.}\ \bibnamefont
  {G{\'o}mez-Urrea}}, \bibinfo {author} {\bibfnamefont {C.~A.}\ \bibnamefont
  {Duque}}, \bibinfo {author} {\bibfnamefont {I.~V.}\ \bibnamefont
  {P{\'e}rez-Quintana}}, \ and\ \bibinfo {author} {\bibfnamefont {M.~E.}\
  \bibnamefont {Mora-Ramos}},\ }\href {\doibase 10.1007/s00340-017-6654-6}
  {\bibfield  {journal} {\bibinfo  {journal} {Appl. Phys. B}\ }\textbf
  {\bibinfo {volume} {123}},\ \bibinfo {pages} {66} (\bibinfo {year}
  {2017})}\BibitemShut {NoStop}%
\bibitem [{\citenamefont {Ramanujam}\ and\ \citenamefont
  {Wilson}(2017)}]{ramanujam_effect_2017}%
  \BibitemOpen
  \bibfield  {author} {\bibinfo {author} {\bibfnamefont {N.~R.}\ \bibnamefont
  {Ramanujam}}\ and\ \bibinfo {author} {\bibfnamefont {K.~S.~J.}\ \bibnamefont
  {Wilson}},\ }\href {\doibase 10.1016/j.optcom.2016.10.059} {\bibfield
  {journal} {\bibinfo  {journal} {Optics Communications}\ }\textbf {\bibinfo
  {volume} {386}},\ \bibinfo {pages} {65} (\bibinfo {year} {2017})}\BibitemShut
  {NoStop}%
\bibitem [{\citenamefont {Askenazi}\ \emph {et~al.}(2014)\citenamefont
  {Askenazi}, \citenamefont {Vasanelli}, \citenamefont {Delteil}, \citenamefont
  {Todorov}, \citenamefont {Andreani}, \citenamefont {Beaudoin}, \citenamefont
  {Sagnes},\ and\ \citenamefont {Sirtori}}]{askenazi_ultra-strong_2014}%
  \BibitemOpen
  \bibfield  {author} {\bibinfo {author} {\bibfnamefont {B.}~\bibnamefont
  {Askenazi}}, \bibinfo {author} {\bibfnamefont {A.}~\bibnamefont {Vasanelli}},
  \bibinfo {author} {\bibfnamefont {A.}~\bibnamefont {Delteil}}, \bibinfo
  {author} {\bibfnamefont {Y.}~\bibnamefont {Todorov}}, \bibinfo {author}
  {\bibfnamefont {L.~C.}\ \bibnamefont {Andreani}}, \bibinfo {author}
  {\bibfnamefont {G.}~\bibnamefont {Beaudoin}}, \bibinfo {author}
  {\bibfnamefont {I.}~\bibnamefont {Sagnes}}, \ and\ \bibinfo {author}
  {\bibfnamefont {C.}~\bibnamefont {Sirtori}},\ }\href {\doibase
  10.1088/1367-2630/16/4/043029} {\bibfield  {journal} {\bibinfo  {journal}
  {New J. Phys.}\ }\textbf {\bibinfo {volume} {16}},\ \bibinfo {pages} {043029}
  (\bibinfo {year} {2014})}\BibitemShut {NoStop}%
\bibitem [{\citenamefont {Wolff}(1970)}]{wolff_plasma-wave_1970}%
  \BibitemOpen
  \bibfield  {author} {\bibinfo {author} {\bibfnamefont {P.~A.}\ \bibnamefont
  {Wolff}},\ }\href {\doibase 10.1103/PhysRevLett.24.266} {\bibfield  {journal}
  {\bibinfo  {journal} {Phys. Rev. Lett.}\ }\textbf {\bibinfo {volume} {24}},\
  \bibinfo {pages} {266} (\bibinfo {year} {1970})}\BibitemShut {NoStop}%
\bibitem [{\citenamefont {Chen}\ and\ \citenamefont
  {Khurgin}(2003)}]{chen_feasibility_2003}%
  \BibitemOpen
  \bibfield  {author} {\bibinfo {author} {\bibfnamefont {J.}~\bibnamefont
  {Chen}}\ and\ \bibinfo {author} {\bibfnamefont {J.~B.}\ \bibnamefont
  {Khurgin}},\ }\href {\doibase 10.1109/JQE.2003.809326} {\bibfield  {journal}
  {\bibinfo  {journal} {IEEE Journal of Quantum Electronics}\ }\textbf
  {\bibinfo {volume} {39}},\ \bibinfo {pages} {600} (\bibinfo {year}
  {2003})}\BibitemShut {NoStop}%
\bibitem [{\citenamefont {Khurgin}\ and\ \citenamefont
  {Liu}(2006)}]{khurgin_stimulated_2006}%
  \BibitemOpen
  \bibfield  {author} {\bibinfo {author} {\bibfnamefont {J.~B.}\ \bibnamefont
  {Khurgin}}\ and\ \bibinfo {author} {\bibfnamefont {H.~C.}\ \bibnamefont
  {Liu}},\ }\href {\doibase 10.1103/PhysRevB.74.035317} {\bibfield  {journal}
  {\bibinfo  {journal} {Phys. Rev. B}\ }\textbf {\bibinfo {volume} {74}},\
  \bibinfo {pages} {035317} (\bibinfo {year} {2006})}\BibitemShut {NoStop}%
\bibitem [{\citenamefont {Kittel}(1963)}]{kittel_quantum_1963}%
  \BibitemOpen
  \bibfield  {author} {\bibinfo {author} {\bibfnamefont {C.}~\bibnamefont
  {Kittel}},\ }\href@noop {} {\emph {\bibinfo {title} {Quantum theory of
  solids}}}\ (\bibinfo  {publisher} {Wiley},\ \bibinfo {address} {New York},\
  \bibinfo {year} {1963})\BibitemShut {NoStop}%
\bibitem [{\citenamefont {Cohen-Tannoudji}\ \emph {et~al.}(1989)\citenamefont
	{Cohen-Tannoudji}, \citenamefont {Dupont-Roc},\ and\ \citenamefont
	{Grynberg}}]{cohen-tannoudji_photons_1989}%
\BibitemOpen
\bibfield  {author} {\bibinfo {author} {\bibfnamefont {C.}~\bibnamefont
		{Cohen-Tannoudji}}, \bibinfo {author} {\bibfnamefont {J.}~\bibnamefont
		{Dupont-Roc}}, \ and\ \bibinfo {author} {\bibfnamefont {G.}~\bibnamefont
		{Grynberg}},\ }\href@noop {} {\emph {\bibinfo
			{title} {Photons and atoms: introduction to quantum electrodynamics}}}\
(\bibinfo  {publisher} {Wiley},\ \bibinfo {address} {New York},\ \bibinfo
{year} {1989})\BibitemShut {NoStop}%
\bibitem [{\citenamefont {Bhatt}\ \emph {et~al.}(1994)\citenamefont {Bhatt},
  \citenamefont {Kim},\ and\ \citenamefont
  {Stroscio}}]{bhatt_theoretical_1994}%
  \BibitemOpen
  \bibfield  {author} {\bibinfo {author} {\bibfnamefont {A.~R.}\ \bibnamefont
  {Bhatt}}, \bibinfo {author} {\bibfnamefont {K.~W.}\ \bibnamefont {Kim}}, \
  and\ \bibinfo {author} {\bibfnamefont {M.~A.}\ \bibnamefont {Stroscio}},\
  }\href {\doibase 10.1063/1.358498} {\bibfield  {journal} {\bibinfo  {journal}
  {J. Appl. Phys.}\ }\textbf {\bibinfo {volume} {76}},\ \bibinfo {pages} {3905}
  (\bibinfo {year} {1994})}\BibitemShut {NoStop}%
\bibitem [{\citenamefont {Wacker}\ \emph {et~al.}(2013)\citenamefont {Wacker},
  \citenamefont {Lindskog},\ and\ \citenamefont
  {Winge}}]{wacker_nonequilibrium_2013}%
  \BibitemOpen
  \bibfield  {author} {\bibinfo {author} {\bibfnamefont {A.}~\bibnamefont
  {Wacker}}, \bibinfo {author} {\bibfnamefont {M.}~\bibnamefont {Lindskog}}, \
  and\ \bibinfo {author} {\bibfnamefont {D.}~\bibnamefont {Winge}},\ }\href
  {\doibase 10.1109/JSTQE.2013.2239613} {\bibfield  {journal} {\bibinfo
  {journal} {Selected Topics in Quantum Electronics, IEEE Journal of}\ }\textbf
  {\bibinfo {volume} {19}},\ \bibinfo {pages} {1200611} (\bibinfo {year}
  {2013})}\BibitemShut {NoStop}%
\bibitem [{\citenamefont {Faist}\ \emph {et~al.}(1994)\citenamefont {Faist},
  \citenamefont {Capasso}, \citenamefont {Sivco}, \citenamefont {Sirtori},
  \citenamefont {Hutchinson},\ and\ \citenamefont {Cho}}]{faist_quantum_1994}%
  \BibitemOpen
  \bibfield  {author} {\bibinfo {author} {\bibfnamefont {J.}~\bibnamefont
  {Faist}}, \bibinfo {author} {\bibfnamefont {F.}~\bibnamefont {Capasso}},
  \bibinfo {author} {\bibfnamefont {D.~L.}\ \bibnamefont {Sivco}}, \bibinfo
  {author} {\bibfnamefont {C.}~\bibnamefont {Sirtori}}, \bibinfo {author}
  {\bibfnamefont {A.~L.}\ \bibnamefont {Hutchinson}}, \ and\ \bibinfo {author}
  {\bibfnamefont {A.~Y.}\ \bibnamefont {Cho}},\ }\href@noop {} {\bibfield
  {journal} {\bibinfo  {journal} {Science}\ }\textbf {\bibinfo {volume}
  {264}},\ \bibinfo {pages} {553} (\bibinfo {year} {1994})}\BibitemShut
  {NoStop}%
\bibitem [{\citenamefont {Ohtani}\ \emph {et~al.}(2016)\citenamefont {Ohtani},
  \citenamefont {Ndebeka-Bandou}, \citenamefont {Bosco}, \citenamefont {Beck},\
  and\ \citenamefont {Faist}}]{ohtani_quantum_2016}%
  \BibitemOpen
  \bibfield  {author} {\bibinfo {author} {\bibfnamefont {K.}~\bibnamefont
  {Ohtani}}, \bibinfo {author} {\bibfnamefont {C.}~\bibnamefont
  {Ndebeka-Bandou}}, \bibinfo {author} {\bibfnamefont {L.}~\bibnamefont
  {Bosco}}, \bibinfo {author} {\bibfnamefont {M.}~\bibnamefont {Beck}}, \ and\
  \bibinfo {author} {\bibfnamefont {J.}~\bibnamefont {Faist}},\ }\href
  {http://arxiv.org/abs/1610.00963} {\bibfield  {journal} {\bibinfo  {journal}
  {arXiv:1610.00963}\ } (\bibinfo {year} {2016})}, \BibitemShut {NoStop}%
\bibitem [{\citenamefont {Terazzi}\ and\ \citenamefont
  {Faist}(2010)}]{terazzi_density_2010}%
  \BibitemOpen
  \bibfield  {author} {\bibinfo {author} {\bibfnamefont {R.}~\bibnamefont
  {Terazzi}}\ and\ \bibinfo {author} {\bibfnamefont {J.}~\bibnamefont
  {Faist}},\ }\href@noop {} {\bibfield  {journal} {\bibinfo  {journal} {New
  Journal of Physics}\ }\textbf {\bibinfo {volume} {12}},\ \bibinfo {pages}
  {033045} (\bibinfo {year} {2010})}\BibitemShut {NoStop}%
\bibitem [{\citenamefont {Henry}\ and\ \citenamefont
  {Hopfield}(1965)}]{henry_raman_1965}%
  \BibitemOpen
  \bibfield  {author} {\bibinfo {author} {\bibfnamefont {C.~H.}\ \bibnamefont
  {Henry}}\ and\ \bibinfo {author} {\bibfnamefont {J.~J.}\ \bibnamefont
  {Hopfield}},\ }\href {\doibase 10.1103/PhysRevLett.15.964} {\bibfield
  {journal} {\bibinfo  {journal} {Phys. Rev. Lett.}\ }\textbf {\bibinfo
  {volume} {15}},\ \bibinfo {pages} {964} (\bibinfo {year} {1965})}\BibitemShut
  {NoStop}%
\end{thebibliography}

%

\end{document}